\documentclass[twocolumn,aps,prl,showpacs,floatfix]{revtex4}
\usepackage{graphicx}
\usepackage{times}
\usepackage{nicefrac}
\usepackage{amsmath}
\usepackage{amsfonts}
\usepackage{amssymb}
\usepackage{amsthm}
\usepackage{epsf}
\usepackage{bm}
\usepackage{times}

\usepackage{dcolumn}

\newcolumntype{.}{D{x}{}{-1}}

\newcommand{\bfn}{\bf n}
\newcommand{\balpha}{{\mbox{\boldmath$\alpha$}}}

\newcommand{\be}{\begin{eqnarray}}
\newcommand{\ee}{\end{eqnarray}}
\newcommand{\la}{\langle}
\newcommand{\ra}{\rangle}

\newcommand{\veps}{\varepsilon}

\newcommand{\Dmatrix}[4]{
        \left(
        \begin{array}{cc}
        #1  & #2   \\
        #3  & #4   \\
        \end{array}
        \right)
        }

\begin{document}

\title{Model operator approach to the Lamb shift calculations
in relativistic many-electron atoms}

\author{V. M. Shabaev}

\affiliation {Department of Physics, St.Petersburg State University,
Ulianovskaya 1, Petrodvorets, St.Petersburg 198504, Russia}

\author{I. I. Tupitsyn}

\affiliation {Department of Physics, St.Petersburg State University,
Ulianovskaya 1, Petrodvorets, St.Petersburg 198504, Russia}

\author{V. A. Yerokhin}

\affiliation{St. Petersburg State Polytechnical University,
             Polytekhnicheskaya 29, St. Petersburg 195251, Russia}

\begin{abstract}

A model operator approach to calculations of the QED corrections
to energy levels in relativistic many-electron atomic systems is developed.
The model Lamb shift operator is represented by a sum of local and
nonlocal  potentials which are defined  using the results
of {\it ab initio} calculations of the diagonal and nondiagonal matrix
elements of the one-loop QED operator
 with H-like wave functions.  
The model operator can be easily included in any calculations based on
the Dirac-Coulomb-Breit Hamiltonian.
Efficiency of the method
is demonstrated by comparison of the model QED operator results
for the Lamb shifts in  many-electron atoms and ions with 
exact QED calculations.

\end{abstract}
\maketitle

\section{Introduction}

A good starting point for the relativistic atomic calculations
is given by the Dirac-Coulomb-Breit (DCB) equation. 
This equation can be solved with a high accuracy by
using either the configuration-interaction
Dirac-Fock (CI-DF) methods \cite{gra70,des75,bra77,ind90,koz01,gla04,tup05}
or the relativistic many-body perturbation theory (RMBPT) methods
\cite{dzu89,blu90,ind92,ynn94,saf99,por09,dzu12}.
In many cases
the precision of these calculations has reached
a level that requires evaluations of
quantum electrodynamics (QED) effects.
To date, the rigorous calculations of the QED effects
in middle- and high-Z systems
are fully restricted to
the 1/Z perturbation theory
 (see, e.g.,
\cite{yer99,art05,sha08,gla11,vol12} and references therein).
The perturbation theory
methods have been also extended to many-electron
ions and atoms employing an effective screening potential
instead of the Coulomb one
\cite{sap02,sap03,sha05a,sha05b,che06,art07,koz10,sap11}.
However, these methods are
too complicated to be directly included
into the DCB calculations.
 For this reason, numerous attempts
have been undertaken to propose simple methods
for incorporating
the QED corrections into
the CI-DF and RMBPT codes. These methods
(see, e.g., \cite{ind90,pyy03,dra03,fla05,thi10,pyy12,tup13,low13,rob13}
and references therein)
are generally based
on scaling the Lamb shift results for the Coulomb
potential to other atomic potentials
which include partially the screening effects.
Such a scaling can be done either directly by using the
Welton's idea \cite{wel48}
to express the main part of the self energy contribution
in terms of $\vec{\nabla}^2 V$ \cite{ind90,dra03,low13}
or by introducing an effective short-range potential
which fits the Lamb shifts for hydrogenlike ions
\cite{pyy03,fla05,thi10,pyy12,tup13,rob13}.

In Ref. \cite{sha93} it was shown that the QED corrections
can be systematically included into an effective Hamiltonian
 acting in the space of the Slater determinants made up of
one-electron positive-energy states whose total (many-electron)
energies are smaller
than the pair-creation threshold. To the lowest order,
this approach leads to a QED operator that, in principle, 
can be added to DCB Hamiltonian. The main goal of this 
paper is to represent this QED operator in a form that can
be easily included in any calculations based on the DCB 
equation.  

In the next section, we summarize the basic equations for the
effective Hamiltonian that includes the one-loop
QED corrections.   Then we approximate
the QED operator by a sum of  short-range local and nonlocal
potentials and calculate the model QED operator in a wide range
of $Z=10-120$. Finally, the model QED operator is applied
to calculations of the Lamb shifts in many-electron
atoms and ions, and the results obtained are compared 
with other QED calculations.

Relativistic units ($\hbar=c=1$) are used in the paper.

\section{Effective Hamiltonian in the framework of QED}

The systematic method to derive a Schr\"odinger-like equation
for a relativistic many-electron atom from QED
can be formulated
within the two-time Green function (TTGF) method \cite{sha02}.
To determine such an equation, first of all one needs
to choose the active space in which the effective Hamiltonian acts.
Since rigorous calculations of the QED effects 
in middle- and high-$Z$ systems
employ  
the perturbation theory starting from
the Dirac equation with a local potential,
the active space is generally considered to be formed
either by a single or by (quasi)degenerate 
unperturbed states. These states are given by the Slater
determinants made up of the solutions
of the Dirac equation with the local potential
considered. However, in Ref. \cite{sha93} 
it was shown that the active space
can be extended
to  all unperturbed states made up of 
one-electron positive-energy states whose total (many-electron)
energies are smaller
than the pair-creation threshold.
Moreover, if the consideration is restricted
to the lowest-order QED terms,
the active
space can be extended beyond the pair-creation
threshold. For simplicity, this extention, having
 no effect on the accuracy 
of the calculations, is considered in the present
paper.
Using the derivations presented in Ref. \cite{sha93},
we summarize below the basic equations that are 
obtained with this choice of the active space.

To simplify the equations, 
we assume that
the active space is formed by the Slater determinants
made up of the positive-energy solutions of 
the Dirac equation with the Coulomb potential
$V_{\rm C}(r)=-\alpha Z_{\rm nuc}(r)/r$ (the effective 
charge $Z_{\rm nuc}(r)$  accounts for the nuclear charge 
distribution):
\begin{eqnarray} \label{Dirac}
h^{\rm D}\psi_n = \veps_n\psi_n\,,
\end{eqnarray}
where 
\begin{eqnarray} \label{Dirac1}
h^{\rm D}=\vec{\alpha} \cdot \vec{p} +  m\beta +  V_{\rm C}(r)
\end{eqnarray}
is the one-electron Dirac Hamiltonian.
We note, however, that all the equations can be easily
adopted to the theory with the active space
formed by the solutions in an effective potential 
$V_{\rm eff}(r)$
(e.g., the Dirac-Hartree, the Kohn-Sham or a local
version of the Dirac-Fock potential) that includes
partly the screening effect. In this case, 
the interaction with
the related counterterm ($V_{\rm C}(r)-V_{\rm eff}(r)$) 
must be included in the total Hamiltonian. 
To construct the desired Hamiltonian, one should first
consider the contribution from the
 one-photon exchange Feynman diagram (Fig. 1).
\begin{figure}
\begin{center}
\includegraphics[angle=0,height=0.20\textheight, width=0.40\textwidth]{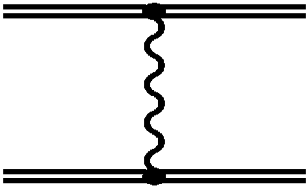}
\caption{\label{fig1} 
One-photon exchange diagram.
}
\end{center}
\end{figure}
Evaluation of this diagram with the TTGF method leads
to the following symmetric 
form of the electron-electron interaction 
operator
(for details see Refs. \cite{sha93,sha02}):
\begin{eqnarray}\label{int}
h^{\rm int} &=&
\sum_{k\ne l,m \ne n}^{(\veps_k,\veps_l,\veps_m,\veps_n>0)}
|\psi_k \psi_l \rangle
\langle \psi_k\psi_l |\frac{1}{2}[I(\varepsilon_{k}-\varepsilon_{m})
\nonumber\\
&& + I(\varepsilon_{l}-\varepsilon_{n})]|\psi_m\psi_n\rangle 
\langle \psi_m \psi_n|\,,
\end{eqnarray}
where the indeces $k$, $l$, $m$, $n$ enumerate the positive-energy one-electron
Dirac states,
$|\psi_k \psi_l  \rangle \equiv |\psi_k\ra | \psi_l  \rangle$
is the direct product of the one-electron Dirac wave functions 
$\psi_k(\vec{r_1})$ and $\psi_l(\vec{r_2})$,
\begin{eqnarray}
I(\omega)=e^2\alpha_1^{\rho}\alpha_2^{\sigma}D_{\rho \sigma}(\omega,r_{12})\,,
\end{eqnarray}
 $\alpha^{\rho}\equiv\gamma^0 \gamma^{\rho}=(1,\balpha)$ are the Dirac matrices, 
$D_{\rho \sigma}(\omega,r_{12})$
is the photon propagator, and $r_{12} = |\vec{r_1}-\vec{r_2}|$
is the interelectronic distance.
It should be noted that 
the symmetric form of the frequency-dependent
electron-electron interaction was first considered in Ref. 
\cite{mit72}.
 The operator (\ref{int}) defines
 the interaction between two electrons only. 
To get the total electron-electron interaction operator for
a many-electron atom, one has to sum Eq. (\ref{int})
over all pairs of atomic electrons:
\begin{eqnarray}\label{int2}
H^{\rm int} = \sum_{i<j} h_{ij}^{\rm int}\,,
\end{eqnarray}
where 
$h_{ij}^{\rm int}$ is the two-electron operator  (\ref{int})
taken for electrons $i$ and $j$.

Taking $D_{\rho \sigma}(\omega,r_{ij})$ in the Coulomb gauge  at zero energy 
transfer  ($\omega=0$)
 leads to the Dirac-Coulomb-Breit Hamiltonian \cite{suc80}: 
\begin{eqnarray}
H^{\rm DCB}=
\Lambda^{(+)}\Bigl[\sum_{i}h_i^{\rm D}  +\sum_{i<j}V_{ij}\Bigr]\Lambda^{(+)}\,,
\end{eqnarray}
where the indeces $i$ and $j$ enumerate the atomic electrons,
$\Lambda^{(+)}$ is the product of the one-electron
projectors on the positive-energy states
(which correspond to the potential $V_{\rm C}$),
$h_i^{\rm D}$ is the one-electron Dirac Hamiltonian
(\ref{Dirac1}) taken for electron $i$,
\begin{eqnarray}
V_{ij} &=& e^2\alpha_i^{\rho}\alpha_j^{\sigma}D_{\rho \sigma}(0,r_{ij})
= V^{\rm C}_{ij}+ V^{\rm B}_{ij}\nonumber \\
&=&\frac{\displaystyle \alpha}{\displaystyle r_{ij}} 
-\alpha\Bigl[\frac{\displaystyle{ \vec{\alpha}_i\cdot \vec{\alpha}_j}}
{\displaystyle{ r_{ij}}}+\frac{\displaystyle 1}{\displaystyle 2}
(\vec{\nabla}_i\cdot \vec{\alpha}_i)(\vec{\nabla}_j\cdot \vec{\alpha}_j)
r_{ij}\Bigr]
\,
\end{eqnarray}
is the the sum of the Coulomb and Breit electron-electron interaction operators,
and $\alpha$ is the fine structure constant.
It is well known that the DCB Hamiltonian
 accounts for the nonrelativistic and lowest-order relativistic 
contributions. 
 In the Feynman gauge, to get the Hamiltonian to the same accuracy,
one has to account for the higher-order photon exchange diagrams (see Ref. \cite{sha93}
and references therein).

As the next step, one should consider the contributions from
the one-loop
self-energy (SE) and vacuum-polarization (VP) diagrams
presented in Figs. 2 and 3, respectively. 
\begin{figure}
\begin{center}
\includegraphics[angle=0,height=0.12\textheight, width=0.30\textwidth]{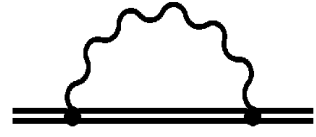}
\caption{\label{fig2} 
Self-energy diagram.
}
\end{center}
\end{figure}
\begin{figure}
\begin{center}
\includegraphics[angle=0,height=0.158\textheight, width=0.17\textwidth]{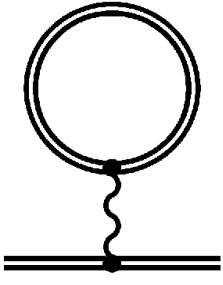}
\caption{\label{fig3} 
Vacuum-polarization diagram.
}
\end{center}
\end{figure}
The direct calculation of these
contributions 
within the TTGF method  
leads to the following symmetric form
of the one-electron QED
 operator $h^{\rm QED}$ \cite{sha93}:
\begin{eqnarray}\label{oper1}
h^{\rm QED}&=&h^{\rm SE}+h^{\rm VP}=  \sum_{k,l}^{(\veps_k,\veps_l >0)}
|\psi_k\ra
\la \psi_k|
\nonumber \\ && \times
\bigl[\frac{1}{2}(\Sigma^{\rm SE}(\veps_k)+\Sigma^{\rm SE}(\veps_l))+V^{\rm VP}\bigr]|\psi_l\ra
\la\psi_l|
\,,  \nonumber \\
\end{eqnarray}
where $\Sigma^{\rm SE}(\veps_k)$ and $ V^{\rm VP}$ are the renormalized SE and VP operators,
respectively, and the sums over  $k$ and $l$ go over all
 the positive-energy one-electron
Dirac states.
To get the total QED operator for a many-electron atom,
one has to sum Eq. (\ref{oper1}) over all 
atomic electrons:
\begin{eqnarray}\label{oper1a}
H^{\rm QED} &=& H^{\rm SE}+ H^{\rm VP}\nonumber\\
&=&\sum_{i} h_i^{\rm QED}
= \sum_{i} (h_i^{\rm SE}+h_i^{\rm VP}) \,,
\end{eqnarray}
where $h_i^{\rm QED}$, 
$h_i^{\rm SE}$, and  $h_i^{\rm VP}$ are the
one-electron operators (\ref{oper1}) taken for electron $i$.

Thus, within the lowest-order QED approximation
 the total effective Hamiltonian can be presented as
\begin{eqnarray} \label{Htot}
H=\Lambda^{(+)}\Bigl[\sum_{i} (h_i^{\rm D}+ h_i^{\rm QED})
+\sum_{i<j} h_{ij}^{\rm int}
\Bigr]  \Lambda^{(+)},
\end{eqnarray}
where  the sums go over all atomic electrons
and  $h_{ij}^{\rm int}$, which  
 is the two-electron operator  (\ref{int})
considered for electrons $i$ and $j$,
 must be taken in the Coulomb gauge
(if it is taken in the Feynman gauge, 
 an additional term must be
added to keep the same accuracy level \cite{sha93}).
In Eq. (\ref{Htot}), all the operators are
 defined for the Coulomb potential $V_{\rm C}$. 
To get the corresponding Hamiltonian
in the active space formed by the solutions in
an effective potential  $V_{\rm eff}$, one should do 
the evident replacements in all the operators, including
the projector, and add the term  
$\Lambda^{(+)}\sum_{i} (V_{\rm C}(r_i)-V_{\rm eff}(r_i))\Lambda^{(+)}$.

\section{Model QED operator}

Since the one-electron VP operator is given by the sum of
the local Uehling and Wichmann-Kroll potentials  \cite{moh98},
$V^{\rm VP} =  V_{\rm Uehl} +  V_{\rm WK}$, it can be easily
included into the DCB equations.
 As is known,
the dominant part
of the VP contribution is represented  by the Uehling
potential:
\begin {eqnarray} \label{uehlexpr}
V_{\rm Uehl}(r)&=&-\alpha Z
\frac{2\alpha}{3\pi}
\int\limits_0^\infty dr'\; 4\pi r'\rho
(r')
\nonumber \\ &&\times
\int\limits_1^\infty dt \;
(1 +\frac{1}{2t^2})
\frac{\sqrt{t^2-1}}{t^{2}}\nonumber \\
&&\times \frac{e^{(-2m|r-r'|t)}-e^{(-2m(r+r')t)}}
{4mrt} \,,\nonumber \\
\end{eqnarray}
where
 $Z\rho(r)$ is the density of
 the nuclear charge distribution
($\int \rho(r) d{\bf r}=1$).
The Uehling potential can easily be evaluated employing the approximate
formulas from Ref. \cite{ful76}. Including the screening effect  into this
potential causes no problem but hardly affects the result (see, e.g.,
Ref. \cite{sha05b}).
Evaluation of the
Wichmann-Kroll potential is a much more difficult problem \cite{sof88,man89,per93}.
However, to a good accuracy, it can be calculated with the help of the approximate
formulas derived in Ref. \cite{fai91}. Therefore,
in what follows, we restrict our consideration to the SE contribution only.

In principle, the operator $H^{\rm SE}$ defined by Eqs. (\ref{oper1})-(\ref{oper1a})
can be added to the DCB Hamiltonian
to account for the lowest-order SE corrections.
In practice, however, due to absence of rather simple
algorithms for evaluation of the SE contributions for 
arbitrary states,
we have to restrict
 $h^{\rm SE}$ to its matrix elements
between a finite number of low-lying one-electron
Dirac-Coulomb (or other effective potential)
wave functions. This restriction
strongly enlarges
the interaction range of the SE operator
and
excludes highly-excited bound and continuum spectrum components
from the active space.
As a result, such a potential
may lead to the SE corrections which strongly deviate from the correct ones.
For instance, in case of H-like ions
 it gives zero results for the states with the principal quantum number
 $n\ge 4$, provided the sums 
in (\ref{oper1}) are limited by $n_k,n_l\le 3$.
 To solve these problems, we modify
the operator   $h^{\rm SE}$ in two steps.

As the first step, to minimize the deviation of
the representation  (\ref{oper1}), restricted to
a finite number of states, from the exact one,
 we separate out a local (with respect to $r$) potential $V^{\rm SE}_{\rm loc}$ 
from the SE operator and
employ formula (\ref{oper1}) for the remaining SE part only.
Such a separation can be justified by the fact that
a dominant part of the nonrelativistic SE operator can be
represented by a local short-range potential \cite{wel48}.
Due to the conservation
of the angular quantum numbers by the one-electron
SE operator, we can choose the local
part to be different for different $\kappa= (-1)^{j+l+1/2}(j+1/2)$.
With this in mind,
we introduce the projector $P_{\kappa}$ which acts only on the angular variables
and is defined by its kernel as
\begin{eqnarray}\label{projector}
&&P_{\kappa}(\bfn, \bfn^{\prime})
=
\nonumber \\
&&\Dmatrix{\sum_{m}\Omega_{\kappa\, m}({\bfn}) \Omega^{\dagger}_{\kappa\, m}({\bfn}^{\prime})}
{\;\;\displaystyle{0}}
{\displaystyle{0}}{\;\; \sum_{m}\Omega_{-\kappa\, m}({\bfn})\Omega^{\dagger}_{-\kappa\, m}({\bfn}^{\prime}) }\,,
\nonumber \\
\end{eqnarray}
where $ \Omega_{\kappa\, m}({\bfn})$ is a spherical spinor.
Then, the local potential is given by
\begin{eqnarray}\label{local1}
V^{\rm SE}_{\rm loc}=\sum_{\kappa}V^{\rm SE}_{{\rm loc},\kappa}P_{\kappa}\,,
\end{eqnarray}
where $V^{\rm SE}_{{\rm loc},\kappa}(r)$ is a short-range radial potential which
can be chosen differently for different values of $\kappa$. 
Strictly speaking, 
 the potential $V^{\rm SE}_{\rm loc}$ is semilocal. However, here and in what follows 
we label it as "local",  keeping in mind that it is local only at a given 
value of $\kappa$.
We put
\begin{eqnarray}\label{local}
 V^{\rm SE}_{{\rm loc},\kappa}(r) =  A_{\kappa} \exp{(-r/\lambdabar_C)}\,,
\end{eqnarray}
where the constant $A_{\kappa}$ is chosen to reproduce
the  SE shift for the lowest energy level at the given $\kappa$
in the corresponding H-like ion, and $\lambdabar_C=\hbar/(mc)$.

As the second step, we restrict the active space of
 the remaining SE operator, $h^{\rm SE}-V^{\rm SE}_{\rm loc}$,
to the basis functions $\{\phi_i({\bf r})\}_{i=1}^{n}$  which, having the same
angular parts as the 
one-electron Dirac-Coulomb
functions $\{\psi_i({\bf r})\}_{i=1}^{n}$,
are localized at smaller distances compared to the
Dirac-Coulomb ones.
The specific choice of the  functions $\{\phi_i({\bf r})\}_{i=1}^{n}$ 
will be given below. 
With these  functions, we approximate the
one-electron SE operator  as follows
\begin{eqnarray}\label{oper2}
h^{\rm SE} = V^{\rm SE}_{\rm loc} + 
\sum_{i,k=1}^{n} |\phi_i\ra B_{ik}\la \phi_k|\,,
\end{eqnarray}
where the matrix $B_{ik}$ has to be determined in such a way
that the matrix elements of the model SE operator (\ref{oper2})
with the H-like wave functions, corresponding to the 
space under consideration, coincide with the exact ones.
This leads to the equations
\begin{eqnarray}\label{oper2a}
\sum_{j,l=1}^{n} \la \psi_i
|\phi_j\ra B_{jl}\la \phi_l|\psi_k\ra
\;\;\;\;\;\;\;\;\;\;\;\;\;\;\;\;\;\;\;\;\;\;\;
\;\;\;\;\;\;\;\;\;\;\;\;\;\;\;\;\;\;\;\;\;\;
 \nonumber \\
=\la \psi_i|\bigl[\frac{1}{2}(\Sigma(\veps_i)+ \Sigma(\veps_k))
-V^{\rm SE}_{\rm loc}\bigr]|\psi_k\ra \,.
\end{eqnarray}
Introducing the matrix  $D_{ik} = \la \phi_i|\psi_k\ra $,
 we get
\begin{eqnarray}
B_{ik}&=&  \sum_{j,l=1}^{n} ((D^t)^{-1})_{ij} 
\la \psi_j|\bigl[\frac{1}{2}(\Sigma(\veps_j)+ \Sigma(\veps_l))
\nonumber\\
&& -V^{\rm SE}_{\rm loc}\bigr]|\psi_l\ra
(D^{-1})_{lk}
\,.
\end{eqnarray}
Therefore, the model one-electron SE operator can be written as
\begin{eqnarray}\label{oper3}
h^{\rm SE}&=&  V^{\rm SE}_{\rm loc}+
\sum_{i,k=1}^n \sum_{j,l=1}^n
|\phi_i\ra
((D^t)^{-1})_{ij}
 \nonumber \\ && \times
\la \psi_j|\bigl[\frac{1}{2}(\Sigma(\veps_j)+\Sigma(\veps_l))-V^{\rm SE}_{\rm loc}
\bigr]|\psi_l\ra]\nonumber\\
&&\times (D^{-1})_{lk}
\la \phi_k|\,.
\nonumber \\
\end{eqnarray}

Now let us consider the choice of  the functions
 $\{\phi_i({\bf r})\}_{i=1}^{n}$. From one side,
since we use the SE matrix elements calculated with
the hydrogenlike wave functions, these functions
should be chosen rather close to the H-like ones.
From the other side, because of a short interaction range of the
SE operator, they should vanish at smaller distances
compared to the H-like wave functions.
With this in mind, we construct them using the H-like
wave functions multiplied with the factor
\begin{eqnarray}
\rho_l(r)=\exp{(-2\alpha Z (r/\lambdabar_C)/(1+l))},
\end{eqnarray}
where $l=|\kappa+1/2|-1/2$ is the orbital angular momentum
of the state under consideration.
The simple choice  $\phi_i({\bf r}) = \rho_{l_i}(r)\psi_i({\bf r})$
fits the goal but, due to a rather similar behaviour of the
 wave functions for different values of the principal quantum number
  at small $r$,
gives a matrix $D$ close to degenerate one and, therefore, leads to a rather
singular matrix $D^{-1}$. For this reason, 
we consider
a slightly different choice. In what follows,
we restrict the basis functions by 
 $ns$, $np_{1/2}$,  $np_{3/2}$, $nd_{3/2}$,
and   $nd_{5/2}$ states with the
principal quantum number  $n\le 3$
for the $s$ states and $n\le 4$ 
for the $p$ and $d$ states,
and put
\begin{eqnarray}
\phi_i({\bf r}) = \frac{1}{2}(I-(-1)^{s_i}\beta)
\rho(r)\psi_i({\bf r})\,,
\end{eqnarray}
where $I$ is the identity matrix,
$\beta$ is the standard Dirac matrix, the index $s_i=n_i-l_i$
enumerates the positive energy states at the given $\kappa$,
and $n_i$ is the principal quantum number. 
With this choice, 
one easily finds
\begin{widetext}
\begin{eqnarray}
&D_{11} =\int_0^\infty dr\, r^2 g_1^2(r)\,\rho_l(r)\,,\;\;\;\;
D_{12} =\int_0^\infty dr\, r^2 g_1(r)\,g_2(r)\,\rho_l(r)\,,\nonumber\\
&D_{13} = D_{31}=\int_0^\infty dr\, r^2 g_1(r)\,g_3(r)\,\rho_l(r)\,,\;\;\;\;
D_{21} =\int_0^\infty dr\, r^2 f_1(r)\,f_2(r)\,\rho_l(r)\,,\nonumber\\
&D_{22} =\int_0^\infty dr\, r^2 f_2^2(r)\,\rho_l(r)\,,\;\;\;\;
D_{23} =\int_0^\infty dr\, r^2 f_2(r)\,f_3(r)\,\rho_l(r)\,,\nonumber\\
& D_{32} =\int_0^\infty dr\, r^2 g_2(r)\,g_3(r)\,\rho_l(r)\,,\;\;\;\;
D_{33} =\int_0^\infty dr\, r^2 g_3^2(r)\,\rho_l(r)\,,
\end{eqnarray}
\end{widetext}
where  $g_i(r)$ and $f_i(r)$ are the upper and lower radial components of the
hydrogenlike wave functions and the index $i$ enumerates
the positive-energy states at the given $\kappa$.
The explicit formulas for the calculation of the inverse
matrix  can be found in the standard textbooks.

Thus, in what follows we use the
model one-electron SE operator given by
\begin{eqnarray} \label{hse_fin}
h^{\rm SE}&=& V^{\rm SE}_{\rm loc}+\frac{1}{4}
\sum_{i,k}
\sum_{j,l}
(I-(-1)^{s_i}\beta)
\rho_{l_i}(r)
|\psi_i\ra\nonumber\\
&&\times ((D^t)^{-1})_{ij}
\la \psi_j|\bigl[\frac{1}{2}(\Sigma(\veps_j)+\Sigma(\veps_l))
-V^{\rm SE}_{\rm loc}\bigr]|\psi_l\ra \nonumber\\
&& \times (D^{-1})_{lk}
\la \psi_k|\rho_{l_k}(r) 
(I-(-1)^{s_k}\beta),
\end{eqnarray}
where the summations run over $ns$ states with
the principal quantum number $n\le 3$ 
and over
 $np_{1/2}$,  $np_{3/2}$  
$nd_{3/2}$, $nd_{5/2}$ states
 with $n\le 4$, 
$\rho_{l_i}(r)=\exp{(-2\alpha Z (r/\lambdabar_C)/(1+l_i))}$,
\begin{eqnarray}
D_{ik} &=& \frac{1}{2} \langle \psi_i| (I-(-1)^{s_i}\beta)
\rho_{l_i}(r)|\psi_k \rangle\,,
\end{eqnarray}
and $V^{\rm SE}_{\rm loc}$ is defined by Eqs. (\ref{local1})-(\ref{local}) .

\section{Matrix elements of the exact self-energy operator}

To complete our construction of the model SE operator, we need
the diagonal and non-diagonal matrix elements of the exact SE operator
$\Sigma(\veps)$ with the hydrogenlike wave functions. Calculations of the SE corrections
reported previously in the literature \cite{moh92a,moh92b,bei98} were performed for the diagonal matrix
elements only. In the present work, we extend these calculations to both the diagonal and
non-diagonal matrix elements of the one-loop SE operator. Our calculation was carried out into two
steps. First, we evaluated the SE matrix element for the point nucleus by using the numerical method
described in detail in Refs.~\cite{yer99a,yerokhin:05:se}. Next, we separately calculated the finite
nuclear size correction, as described in Ref.~\cite{yerokhin:11:fns}. The finite nuclear size
effect was calculated with the standard two-parameter Fermi model for the nuclear charge distribution.

The results of our calculations for the $ns$, $np_{1/2}$,
$np_{3/2}$, $nd_{3/2}$, and $nd_{5/2}$ states
with $n$ up to 5 are presented in Tables~\ref{tab:s}, \ref{tab:p1},
\ref{tab:p3}, \ref{tab:d3}, \ref{tab:d5}, respectively.
They are expressed in terms of the function $F_{ik}(\alpha Z)$
defined by
\begin{eqnarray}\label{F}
\Sigma_{ik}&\equiv&
\la \psi_i|\frac{1}{2}(\Sigma(\veps_i)+\Sigma(\veps_k))|\psi_k\ra
 \nonumber \\
&=&\frac{\alpha}{\pi}\frac{(\alpha Z)^4}{(n_i n_k)^{3/2}}
F_{ik}(\alpha Z)mc^2\,,
\end{eqnarray}
where $n_i$ and $n_k$ are the principal quantum numbers
of the $i$ and $k$ states, respectively. In the tables, the results are presented separately for
the point nucleus and for the extended nucleus (except for the cases when both results coincide).
If no error is specified, the results are supposed to be accurate to all digits quoted. In the case
of diagonal matrix elements, excellent agreement with previous results \cite{moh92a,moh92b,bei98}
is observed. In this paper, to define the model SE operator we use
 the values $F_{ik}(\alpha Z)$ with the principal quantum numbers
$n_i, n_k \le 3$  for the $s$ states and $n_i, n_k \le 4$  for the $p$ and $d$ states.
As to the other data presented in the tables, they can be used
for test calculations with H-like ions 
(see the next section) as well as for extending the active space
of the model SE operator.

To obtain  the function $F_{ik}(\alpha Z)$ for values of $Z$ not listed in the tables, one may use
a polynomial interpolation, applied to the function
\begin{eqnarray}
G_{ik}(\alpha Z) = F_{ik}(\alpha Z)-\delta_{l0}(4/3){\rm ln}(\alpha Z)^{-2}\,.
\end{eqnarray}
Here, following to Ref. \cite{moh83},
we have subtracted the log term which represents the small-$\alpha Z$ behaviour
of  $F_{ik}(\alpha Z)$ for $s$ states ($l=0$).
The interpolation function is thus given by \cite{moh83}
\begin{eqnarray}
F'_{ik}(\alpha Z)&=&
\delta_{l0}(4/3){\rm ln}(\alpha Z)^{-2}
+\sum_{n=1}^{N}\Bigl[\prod_{m\ne n}\frac{Z-Z_m}{Z_n-Z_m}\Bigr]
 \nonumber \\ && \times
[F_{ik}(\alpha Z_n)-\delta_{l0}(4/3){\rm ln}(\alpha Z_n)^{-2}]\,.
\end{eqnarray}

\section{Calculations with the model self-energy operator}

Since the model SE operator is constructed
using the SE matrix elements with H-like wave functions
of the $ns$ states at $n\le 3$ and 
the $np$ and $nd$ states at $n\le 4$, first of all, we 
should consider how it works for H-like states 
 with higher values of $n$. In  Table~\ref{tab:H-like}
we present the SE shifts 
for 
 the $4s$, $5s$, $5p_{1/2}$, $5p_{3/2}$, $5d_{3/2}$,
and $5d_{5/2}$ states in
H-like ions,
 obtained using the model SE operator, $\la v|h^{\rm SE}|v\ra $, 
and compare them with the corresponding exact 
results. 
To demonstrate the importance of the nonlocal
part of $h^{\rm SE}$, 
we present also the local  $\la v|V_{\rm loc}^{\rm SE}|v\ra$
contribution. As one can see from the table, for the  $s$ states 
the difference between the exact and model SE operator results
does not exceed 1\%. As to the $p$ and $d$ states, despite the relative
value of the difference is significantly bigger than for the $s$ states,
its absolute value,
expressed in terms of the functions $F(\alpha Z)$, does not exceed
0.01.
We stress also the importance of the nonlocal part of the
model SE operator: for the $s$ states the difference between
the local part and the total result can amount to about 30\%.  

To demonstrate the efficiency of the method, we also applied it
to calculations of the Lamb shifts in
neutral alkali metals,
Cu-like ions, superheavy atoms, and Li-like ions.

Calculations of the Lamb shift
in alkali metals were considered in Refs. \cite{sap02,lab99}.
In Ref. \cite{sap02}, it was calculated in
the potential $V_{\rm eff}(r)$, which is defined in terms of an effective
charge $Z_{\rm eff}(r)$ through
\begin{eqnarray}\label{U}
V_{\rm eff}(r) =-\frac{\alpha Z_{\rm eff}(r)}{r}\,,
\end{eqnarray}
where
\begin{eqnarray}
 Z_{\rm eff}(r)=Z_{\rm nuc}(r) -r\int_0^{\infty}dr'\frac{1}{r_{>}}
\rho_{\rm t}(r')+x_{\alpha}\Bigl[\frac{81}{32\pi^2}r\rho_{\rm t}\Bigr]^{1/3}
\end{eqnarray}
and $\rho_{\rm t}=\rho_{\rm v}+\rho_{\rm c}$ is total (valence plus core)
electron charge density,
which is determined self-consistently solving the Dirac equation with the
potential $V_{\rm eff}(r)$ (see Ref. \cite{sap02} for details).
The choice $x_{\alpha}=0$ corresponds to the Dirac-Hartree potential,
 $x_{\alpha}= 2/3$  gives the Kohn-Sham potential, and $x_{\alpha}= 1$
is the Dirac-Slater potential.
In Table~\ref{tab:alkali}, we present the results of our calculations
of the SE contribution to the Lamb
shift performed for $x_{\alpha}=0,\, 1/3,\, 2/3,\, 1$,
and related exact
data by Sapirstein and Cheng  \cite{sap02}.
Our data were obtained by averaging the model SE operator
$h^{\rm SE}$, given by Eq. (\ref{hse_fin}),
with the wave function of the valence state $v$ determined from the Dirac
equation with the potential $V_{\rm eff}(r)$. To demonstrate the importance of the nonlocal
part of $h^{\rm SE}$, in addition to the total $\la v|h^{\rm SE}|v\ra $ contribution,
we present also the local  $\la v|V_{\rm loc}^{\rm SE}|v\ra$
part. As one can see from the table,
 for all atoms the $\la v|h^{\rm SE}|v\ra $ values are in a good agreement
with the exact results, while the local potential
approximation,
 $\la v|V_{\rm loc}^{\rm SE}|v\ra$, works reasonably well only
for low-$Z$ systems.

In Ref. \cite{che06}, the QED corrections  to the transition energies
in Cu-like ions have been calculated in the Kohn-Sham potential. In Table~\ref{tab:cu},
we present the SE corrections to the $4s-4p_{1/2}$, $4s-4p_{3/2}$,
  $4p_{1/2}-4d_{3/2}$, $4p_{3/2}-4d_{3/2}$, and $4p_{3/2}-4d_{5/2}$
transition energies
 obtained by averaging the
model SE operator with the wave function of the valence electron.
This wave function was calculated by solving the KS equation
with the KS potential constructed self-consistently with the 4s state.
The comparison with the related exact results from Ref. \cite{che06}
is also given. It can be seen that the model SE operator results are
in a very good agreement with the exact results.

Experiments with superheavy elements have triggered a great interest
to calculations of the QED effects in superheavy atoms
\cite{pyy12,lab99,ind07,goi09,thi10}. In Ref. \cite{goi09},
the QED contributions to the binding energy
of the valence $7s$ electrons in Rg ($Z=111$) and Cn ($Z=112$)
have been evaluated in a local Dirac-Fock potential.
In Table~\ref{tab:comp}, we compare the related SE contributions
obtained using the model SE operator in the
Dirac-Fock method
with those
by Goidenko   \cite{goi09}. Despite the calculations
in the local DF potential
are not fully equivalent to the
calculations based on the DF equations,
 the SE contributions to the one-electron
binding energies obtained by averaging the model SE operator
with the DF wave function of the valence electron as well as
by including
this operator into the DF equations
are in a good agreement with the corresponding
Goidenko's results. 
For comparison, we present also the total DF values which are 
obtained as the difference between the SE contributions to 
the total DF energies of the atom and the ion. 
It is known \cite{eli94,eli95}
that for Rg and Cn the ionization
occurs out of the $6d_{5/2}$ shell instead of the $7s$ shell. 
However, for comparison purposes, here we consider
 the  $^2S_{1/2}\rightarrow ^1S_0$ transition for Rg and 
the $^1S_{0}\rightarrow ^2S_{1/2}$ transition for Cn.
The presented results are also in a reasonable agreement
 with those
based on the Welton method \cite{ind07}
and with the results obtained using a local SE
potential \cite{thi10}.
 In Table~\ref{tab:comp}, we give also the results of our
calculations of the SE corrections to the  binding energy
of the valence $8s$ electrons
in E119 and E120 and compare them
with the values obtained in Ref. \cite{thi10}.

Finally, we applied our model approach to the Li-like ions, 
for which rigorous QED calculations have been performed.
The self-energy screening diagrams
for Li-like ions to first order in $1/Z$
were first evaluated in Ref. \cite{yer99}.
In that paper, the calculations were performed in the Coulomb potential.
 Later, the same diagrams have been calculated in the Kohn-Sham
and core-Hartree potentials
\cite{koz10,sap11}. In Table~\ref{tab:li}, we present the results of our calculations of
the screened SE corrections in Li-like ions,
based on the model SE operator approach,
and compare them with the related results from Refs. \cite{koz10,sap11}.
In our approach, the screened SE corrections  were obtained
by calculating the total
ion energy with the model SE operator included into the Dirac-Fock (DF)
or the  Kohn-Sham (KS)
equation and subtracting both the related energy evaluated
without the model SE operator and the
SE contribution evaluated with the H-like wave functions.
In case of the KS method,  the KS potential
was constructed self-consistently with the valence state
under consideration.
As one can see from the table, the model SE operator
results obtained employing the KS and DF equations
 are  in a fair agreement with
the results obtained by the perturbation theory
 \cite{koz10,sap11}.
Therefore,  to a good accuracy,
the total SE corrections
can be obtained by
summing  the H-like
SE contributions and
the screened SE corrections evaluated by solving
either DF or KS equations with the model SE operator
included, as described above.
\section{Conclusion}
In this paper we have developed the model QED operator approach
to calculations of the Lamb shifts in relativistic atomic systems. 
With this method, we proposed the model self-energy operator
which is given by Eq. (\ref{hse_fin}). This operator can be easily
incorporated into any calculations employing the Dirac-Coulomb-Breit
Hamiltonian. This was demonstrated by calculating the Lamb shifts
in atoms and ions with the use of the model SE operator and comparing
the obtained results with corresponding exact QED calculations.


\section{Acknowledgments}
We thank Andrey Volotka 
for providing
us with details of his calculations of the self-energy screening diagrams
performed in Ref. \cite{koz10}. Valuable discussions with Jan Derezinski,
Bogumil Jeziorski,  
and Jonathan Sapirstein are gratefully acknowledged.
This work was supported by RFBR (Grants No. 13-02-00630   and  No.11-02-00943-a)
and by the Ministry of Education and Science of Russian Federation
(Grant No. 8420).
%
%
\begingroup
\squeezetable
\begin{table*}
\caption{Self-energy correction for $ns$ states. Labelling $(n,n')$ denotes the
  $F_{nn'}$ function defined by Eq.~(\ref{F}). $R$ is the root-mean-square
  charge radius of the nucleus (in fermi). For each $Z$, the upper line
  shows the point-nucleus result, whereas the lower line displays the
  extended-nucleus result (if different).
\label{tab:s}}
\begin{ruledtabular}
\begin{tabular}{lclllllllllllllll}
 $Z$ & $R$     &(1,1)                           &(1,2)                           &(1,3)                           &(1,4)                           &(1,5)                           &(2,2)                           &(2,3)                           &(2,4)                           &(2,5)                           &(3,3)                           &(3,4)                           &(3,5)                           &(4,4)                           &(4,5)                           &(5,5)                         \\ \colrule\\[-5pt]
 10 & 3.005  &   4.6542    &   4.7961    &   4.8145    &   4.8193    &   4.8210(1) &   4.8944    &   4.9325    &   4.9437    &   4.9480    &   4.9524    &   4.9677    &   4.9740    &   4.9749    &   4.9824    &   4.9858     \\[2pt]
 15 & 3.189  &   3.8014    &   3.9463    &   3.9639    &   3.9681    &   3.9693    &   4.0509    &   4.0885    &   4.0992    &   4.1030    &   4.1082    &   4.1229    &   4.1288    &   4.1296    &   4.1367    &   4.1396     \\
    &        &   3.8013    &   3.9462    &   3.9639    &   3.9680    &   3.9693    &   4.0508    &   4.0885    &   4.0991    &   4.1030    &   4.1082    &   4.1229    &   4.1287    &   4.1296    &   4.1366    &   4.1396     \\[2pt]
 20 & 3.476  &   3.2463    &   3.3946    &   3.4114    &   3.4148    &   3.4155    &   3.5066    &   3.5438    &   3.5538    &   3.5572    &   3.5633    &   3.5773    &   3.5826    &   3.5834    &   3.5899    &   3.5923     \\
    &        &   3.2462    &   3.3945    &   3.4113    &   3.4147    &   3.4154    &   3.5065    &   3.5437    &   3.5537    &   3.5571    &   3.5632    &   3.5772    &   3.5825    &   3.5833    &   3.5898    &   3.5922     \\[2pt]
 25 & 3.706  &   2.8501    &   3.0023    &   3.0183    &   3.0208    &   3.0209    &   3.1230    &   3.1597    &   3.1689    &   3.1716    &   3.1789    &   3.1922    &   3.1969    &   3.1975    &   3.2034    &   3.2052     \\
    &        &   2.8499    &   3.0022    &   3.0182    &   3.0207    &   3.0207    &   3.1228    &   3.1595    &   3.1688    &   3.1715    &   3.1787    &   3.1920    &   3.1967    &   3.1974    &   3.2032    &   3.2050     \\[2pt]
 30 & 3.929  &   2.5520    &   2.7087    &   2.7238    &   2.7253    &   2.7246    &   2.8388    &   2.8750    &   2.8835    &   2.8855    &   2.8940    &   2.9064    &   2.9105    &   2.9110    &   2.9162    &   2.9173     \\
    &        &   2.5518    &   2.7084    &   2.7235    &   2.7251    &   2.7244    &   2.8386    &   2.8748    &   2.8832    &   2.8853    &   2.8937    &   2.9062    &   2.9102    &   2.9108    &   2.9159    &   2.9171     \\[2pt]
 35 & 4.163  &   2.3200    &   2.4816    &   2.4958    &   2.4963    &   2.4949    &   2.6223    &   2.6580    &   2.6656    &   2.6669    &   2.6767    &   2.6882    &   2.6915    &   2.6919    &   2.6963    &   2.6967     \\
    &        &   2.3196    &   2.4812    &   2.4954    &   2.4960    &   2.4945    &   2.6220    &   2.6576    &   2.6652    &   2.6665    &   2.6763    &   2.6878    &   2.6911    &   2.6915    &   2.6959    &   2.6963     \\[2pt]
 40 & 4.270  &   2.1352    &   2.3024    &   2.3156    &   2.3151    &   2.3129    &   2.4548    &   2.4899    &   2.4965    &   2.4970    &   2.5083    &   2.5187    &   2.5213    &   2.5215    &   2.5250    &   2.5246     \\
    &        &   2.1347    &   2.3019    &   2.3151    &   2.3146    &   2.3124    &   2.4543    &   2.4893    &   2.4960    &   2.4965    &   2.5078    &   2.5182    &   2.5208    &   2.5210    &   2.5245    &   2.5241     \\[2pt]
 45 & 4.494  &   1.9859    &   2.1594    &   2.1716    &   2.1700    &   2.1668    &   2.3247    &   2.3591    &   2.3647    &   2.3643    &   2.3772    &   2.3865    &   2.3882    &   2.3882    &   2.3907    &   2.3894     \\
    &        &   1.9853    &   2.1587    &   2.1709    &   2.1692    &   2.1661    &   2.3239    &   2.3583    &   2.3639    &   2.3635    &   2.3764    &   2.3857    &   2.3874    &   2.3874    &   2.3900    &   2.3887     \\[2pt]
 50 & 4.654  &   1.8643    &   2.0448    &   2.0560    &   2.0531    &   2.0490    &   2.2243    &   2.2580    &   2.2625    &   2.2611    &   2.2757    &   2.2837    &   2.2844    &   2.2842    &   2.2857    &   2.2833     \\
    &        &   1.8633    &   2.0438    &   2.0550    &   2.0521    &   2.0480    &   2.2233    &   2.2569    &   2.2614    &   2.2600    &   2.2747    &   2.2827    &   2.2834    &   2.2832    &   2.2846    &   2.2823     \\[2pt]
 55 & 4.804  &   1.7648    &   1.9533    &   1.9635    &   1.9593    &   1.9542    &   2.1487    &   2.1816    &   2.1847    &   2.1822    &   2.1988    &   2.2054    &   2.2050    &   2.2045    &   2.2047    &   2.2012     \\
    &        &   1.7635    &   1.9519    &   1.9621    &   1.9579    &   1.9527    &   2.1472    &   2.1800    &   2.1832    &   2.1807    &   2.1973    &   2.2039    &   2.2035    &   2.2030    &   2.2032    &   2.1997     \\[2pt]
 60 & 4.912  &   1.6838    &   1.8815    &   1.8906    &   1.8848    &   1.8786    &   2.0945    &   2.1264    &   2.1281    &   2.1243    &   2.1431    &   2.1479    &   2.1463    &   2.1455    &   2.1443    &   2.1395     \\
    &        &   1.6820    &   1.8795    &   1.8886    &   1.8829    &   1.8766    &   2.0923    &   2.1242    &   2.1259    &   2.1222    &   2.1410    &   2.1458    &   2.1442    &   2.1434    &   2.1422    &   2.1374     \\[2pt]
 65 & 5.060  &   1.6186    &   1.8267    &   1.8346    &   1.8273    &   1.8197    &   2.0596    &   2.0904    &   2.0904    &   2.0852    &   2.1064    &   2.1092    &   2.1061    &   2.1049    &   2.1021    &   2.0958     \\
    &        &   1.6161    &   1.8239    &   1.8318    &   1.8245    &   1.8169    &   2.0564    &   2.0872    &   2.0872    &   2.0821    &   2.1033    &   2.1061    &   2.1031    &   2.1019    &   2.0991    &   2.0928     \\[2pt]
 70 & 5.311  &   1.5674    &   1.7876    &   1.7942    &   1.7850    &   1.7760    &   2.0429    &   2.0723    &   2.0703    &   2.0634    &   2.0874    &   2.0878    &   2.0831    &   2.0814    &   2.0767    &   2.0686     \\
    &        &   1.5637    &   1.7833    &   1.7900    &   1.7808    &   1.7719    &   2.0381    &   2.0675    &   2.0655    &   2.0587    &   2.0826    &   2.0832    &   2.0785    &   2.0768    &   2.0721    &   2.0641     \\[2pt]
 75 & 5.339  &   1.5290    &   1.7632    &   1.7683    &   1.7570    &   1.7464    &   2.0441    &   2.0718    &   2.0674    &   2.0586    &   2.0857    &   2.0834    &   2.0767    &   2.0744    &   2.0674    &   2.0573     \\
    &        &   1.5239    &   1.7572    &   1.7624    &   1.7512    &   1.7406    &   2.0373    &   2.0650    &   2.0607    &   2.0519    &   2.0790    &   2.0767    &   2.0701    &   2.0678    &   2.0610    &   2.0509     \\[2pt]
 80 & 5.463  &   1.5028    &   1.7533    &   1.7568    &   1.7431    &   1.7305    &   2.0639    &   2.0894    &   2.0822    &   2.0710    &   2.1018    &   2.0961    &   2.0870    &   2.0840    &   2.0744    &   2.0619     \\
    &        &   1.4955    &   1.7447    &   1.7482    &   1.7346    &   1.7221    &   2.0537    &   2.0793    &   2.0722    &   2.0611    &   2.0917    &   2.0862    &   2.0772    &   2.0743    &   2.0648    &   2.0524     \\[2pt]
 85 & 5.539  &   1.4888    &   1.7587    &   1.7602    &   1.7435    &   1.7287    &   2.1039    &   2.1266    &   2.1158    &   2.1018    &   2.1369    &   2.1271    &   2.1152    &   2.1113    &   2.0985    &   2.0831     \\
    &        &   1.4784    &   1.7461    &   1.7477    &   1.7312    &   1.7165    &   2.0886    &   2.1115    &   2.1009    &   2.0871    &   2.1219    &   2.1124    &   2.1006    &   2.0968    &   2.0842    &   2.0689     \\[2pt]
 90 & 5.710  &   1.4875    &   1.7806    &   1.7797    &   1.7595    &   1.7420    &   2.1669    &   2.1860    &   2.1708    &   2.1533    &   2.1936    &   2.1786    &   2.1631    &   2.1582    &   2.1415    &   2.1225     \\
    &        &   1.4721    &   1.7615    &   1.7607    &   1.7408    &   1.7236    &   2.1431    &   2.1625    &   2.1477    &   2.1305    &   2.1702    &   2.1557    &   2.1406    &   2.1357    &   2.1194    &   2.1007     \\[2pt]
 95 & 5.905  &   1.5005    &   1.8218    &   1.8176    &   1.7931    &   1.7723    &   2.2575    &   2.2719    &   2.2510    &   2.2291    &   2.2757    &   2.2543    &   2.2344    &   2.2280    &   2.2066    &   2.1831     \\
    &        &   1.4772    &   1.7921    &   1.7882    &   1.7642    &   1.7439    &   2.2197    &   2.2345    &   2.2143    &   2.1929    &   2.2386    &   2.2180    &   2.1987    &   2.1925(1) &   2.1717    &   2.1488     \\[2pt]
100 & 5.857  &   1.5302    &   1.8863    &   1.8779    &   1.8479    &   1.8230    &   2.3831    &   2.3910    &   2.3627    &   2.3351    &   2.3897    &   2.3600    &   2.3343    &   2.3262    &   2.2987    &   2.2697     \\
    &        &   1.4961    &   1.8417    &   1.8338    &   1.8047    &   1.7806    &   2.3247    &   2.3332    &   2.3061    &   2.2795    &   2.3325    &   2.3042    &   2.2795    &   2.2717(1) &   2.2453(1) &   2.2172     \\[2pt]
105 & 5.919  &   1.5809    &   1.9811    &   1.9670    &   1.9298    &   1.8996    &   2.5553    &   2.5543    &   2.5158    &   2.4807    &   2.5457    &   2.5051    &   2.4718    &   2.4615    &   2.4261    &   2.3899     \\
    &        &   1.5286    &   1.9107    &   1.8974    &   1.8619    &   1.8331    &   2.4602    &   2.4604    &   2.4243    &   2.3911    &   2.4531    &   2.4150    &   2.3836    &   2.3737    &   2.3403(1) &   2.3058     \\[2pt]
110 & 5.993  &   1.6601    &   2.1179    &   2.0956    &   2.0487    &   2.0114    &   2.7936    &   2.7797    &   2.7272    &   2.6819    &   2.7609    &   2.7055    &   2.6617    &   2.6484    &   2.6025    &   2.5566     \\
    &        &   1.5771    &   2.0031    &   1.9824    &   1.9388    &   1.9040    &   2.6334    &   2.6222    &   2.5743    &   2.5325    &   2.6058    &   2.5552    &   2.5151    &   2.5026(1) &   2.4604(2) &   2.4179     \\[2pt]
115 & 6.088  &   1.7811    &   2.3180    &   2.2832    &   2.2224    &   2.1751    &   3.1331    &   3.1001    &   3.0270    &   2.9668    &   3.0659    &   2.9893    &   2.9307    &   2.9131    &   2.8527    &   2.7933     \\
    &        &   1.6441    &   2.1228    &   2.0917    &   2.0375    &   1.9952    &   2.8519    &   2.8248    &   2.7610    &   2.7081    &   2.7958    &   2.7288    &   2.6775    &   2.6617(1) &   2.6084(1) &   2.5556     \\[2pt]
120 & 6.175  &   1.9709    &   2.6233    &   2.5685    &   2.4866(1) &   2.4243(1) &   3.6437    &   3.5804(1) &   3.4750(1) &   3.3921(1) &   3.5219(1) &   3.4131    &   3.3319    &   3.3080    &   3.2259    &   3.1464     \\
    &        &   1.7335(1) &   2.2753(1) &   2.2294(1) &   2.1613(1) &   2.1093(1) &   3.1256(1) &   3.0760(1) &   2.9909(1) &   2.9231(1) &   3.0295(1) &   2.9411(1) &   2.8750(1) &   2.8549(2) &   2.7875(1) &   2.7216     \\[2pt]
\end{tabular}
\end{ruledtabular}
\end{table*}
\endgroup

%
%
%
\begingroup
\squeezetable
\begin{table*}
\caption{Self-energy correction for $np_{1/2}$ states.
\label{tab:p1}}
\begin{ruledtabular}
\begin{tabular}{ll..........}
 $Z$ & $R$     &\multicolumn{1}{c}{(2,2)}       &\multicolumn{1}{c}{(2,3)}       &\multicolumn{1}{c}{(2,4)}       &\multicolumn{1}{c}{(2,5)}       &\multicolumn{1}{c}{(3,3)}       &\multicolumn{1}{c}{(3,4)}       &\multicolumn{1}{c}{(3,5)}       &\multicolumn{1}{c}{(4,4)}       &\multicolumn{1}{c}{(4,5)}       &\multicolumn{1}{c}{(5,5)}     \\ \colrule\\[-5pt]
 10 & 3.005  &  -0.11x48   &  -0.09x62   &  -0.09x46   &  -0.09x46   &  -0.10x20   &  -0.09x41   &  -0.09x23   &  -0.09x64(1)&  -0.09x21   &  -0.09x32    \\[2pt]
 15 & 3.189  &  -0.10x45   &  -0.08x51   &  -0.08x33   &  -0.08x31   &  -0.09x01   &  -0.08x18   &  -0.07x99   &  -0.08x37   &  -0.07x94   &  -0.08x03    \\[2pt]
 20 & 3.476  &  -0.09x25   &  -0.07x21   &  -0.06x99   &  -0.06x96   &  -0.07x60   &  -0.06x74   &  -0.06x53   &  -0.06x90   &  -0.06x44   &  -0.06x53    \\[2pt]
 25 & 3.706  &  -0.07x91   &  -0.05x76   &  -0.05x50   &  -0.05x45   &  -0.06x03   &  -0.05x12   &  -0.04x89   &  -0.05x25   &  -0.04x77   &  -0.04x84    \\[2pt]
 30 & 3.929  &  -0.06x43   &  -0.04x16   &  -0.03x86   &  -0.03x79   &  -0.04x31   &  -0.03x36   &  -0.03x10   &  -0.03x44   &  -0.02x95   &  -0.02x99    \\[2pt]
 35 & 4.163  &  -0.04x83   &  -0.02x42   &  -0.02x07   &  -0.01x99   &  -0.02x44   &  -0.01x44   &  -0.01x16   &  -0.01x48   &  -0.00x97   &  -0.00x99    \\[2pt]
 40 & 4.270  &  -0.03x10   &  -0.00x54   &  -0.00x14   &  -0.00x03   &  -0.00x42   &   0.00x63   &   0.00x94   &   0.00x63   &   0.01x17   &   0.01x16    \\[2pt]
 45 & 4.494  &  -0.01x23   &   0.01x50   &   0.01x96   &   0.02x08   &   0.01x77   &   0.02x87   &   0.03x21   &   0.02x91   &   0.03x47   &   0.03x47    \\[2pt]
 50 & 4.654  &   0.00x80   &   0.03x71   &   0.04x23   &   0.04x38   &   0.04x14   &   0.05x29   &   0.05x66   &   0.05x38   &   0.05x95   &   0.05x97    \\[2pt]
 55 & 4.804  &   0.03x03   &   0.06x14   &   0.06x72   &   0.06x89   &   0.06x72   &   0.07x93   &   0.08x33   &   0.08x05   &   0.08x64   &   0.08x67    \\[2pt]
 60 & 4.912  &   0.05x48   &   0.08x81   &   0.09x46   &   0.09x65   &   0.09x56   &   0.10x83   &   0.11x25   &   0.10x98   &   0.11x58   &   0.11x61    \\
    &        &   0.05x47   &   0.08x81   &   0.09x46   &   0.09x65   &   0.09x55   &   0.10x82   &   0.11x25   &   0.10x98   &   0.11x57   &   0.11x61    \\[2pt]
 65 & 5.060  &   0.08x20   &   0.11x78   &   0.12x50   &   0.12x71   &   0.12x71   &   0.14x02   &   0.14x48   &   0.14x21   &   0.14x81   &   0.14x85    \\
    &        &   0.08x19   &   0.11x77   &   0.12x49   &   0.12x70   &   0.12x70   &   0.14x01   &   0.14x47   &   0.14x20   &   0.14x80   &   0.14x84    \\[2pt]
 70 & 5.311  &   0.11x27   &   0.15x12   &   0.15x91   &   0.16x14   &   0.16x23   &   0.17x60   &   0.18x07   &   0.17x81   &   0.18x40   &   0.18x44    \\
    &        &   0.11x26   &   0.15x10   &   0.15x90   &   0.16x12   &   0.16x21   &   0.17x58   &   0.18x05   &   0.17x78   &   0.18x38   &   0.18x41    \\[2pt]
 75 & 5.339  &   0.14x77   &   0.18x91   &   0.19x79   &   0.20x02   &   0.20x23   &   0.21x64   &   0.22x13   &   0.21x86   &   0.22x44   &   0.22x46    \\
    &        &   0.14x74   &   0.18x88   &   0.19x75   &   0.19x99   &   0.20x20   &   0.21x60   &   0.22x09   &   0.21x82   &   0.22x40   &   0.22x42    \\[2pt]
 80 & 5.463  &   0.18x82   &   0.23x29   &   0.24x24   &   0.24x48   &   0.24x83   &   0.26x26   &   0.26x77   &   0.26x49   &   0.27x04   &   0.27x03    \\
    &        &   0.18x77   &   0.23x23   &   0.24x18   &   0.24x41   &   0.24x76   &   0.26x19   &   0.26x70   &   0.26x42   &   0.26x97   &   0.26x96    \\[2pt]
 85 & 5.539  &   0.23x59   &   0.28x43   &   0.29x44   &   0.29x67   &   0.30x20   &   0.31x65   &   0.32x15   &   0.31x85   &   0.32x36   &   0.32x30    \\
    &        &   0.23x49   &   0.28x32   &   0.29x33   &   0.29x55   &   0.30x08   &   0.31x52   &   0.32x02   &   0.31x73   &   0.32x23   &   0.32x17    \\[2pt]
 90 & 5.710  &   0.29x31   &   0.34x57   &   0.35x63   &   0.35x83   &   0.36x60   &   0.38x03   &   0.38x50   &   0.38x19   &   0.38x62   &   0.38x49    \\
    &        &   0.29x12   &   0.34x36   &   0.35x42   &   0.35x62   &   0.36x37   &   0.37x80   &   0.38x27   &   0.37x95   &   0.38x38   &   0.38x25    \\[2pt]
 95 & 5.905  &   0.36x33   &   0.42x05   &   0.43x15   &   0.43x29   &   0.44x38   &   0.45x75   &   0.46x16   &   0.45x83   &   0.46x13   &   0.45x89    \\
    &        &   0.35x97   &   0.41x66   &   0.42x75   &   0.42x88   &   0.43x94   &   0.45x31   &   0.45x72   &   0.45x38   &   0.45x69   &   0.45x44    \\[2pt]
100 & 5.857  &   0.45x17   &   0.51x43   &   0.52x51   &   0.52x54   &   0.54x08   &   0.55x33   &   0.55x63   &   0.55x25   &   0.55x38   &   0.54x98    \\
    &        &   0.44x50   &   0.50x70   &   0.51x77   &   0.51x80   &   0.53x27   &   0.54x52   &   0.54x82   &   0.54x43   &   0.54x56   &   0.54x16    \\[2pt]
105 & 5.919  &   0.56x69   &   0.63x55   &   0.64x54   &   0.64x38   &   0.66x55   &   0.67x58   &   0.67x68   &   0.67x25   &   0.67x11   &   0.66x45    \\
    &        &   0.55x37   &   0.62x11   &   0.63x08   &   0.62x94   &   0.64x98   &   0.66x00   &   0.66x10   &   0.65x66   &   0.65x52   &   0.64x87    \\[2pt]
110 & 5.993  &   0.72x31   &   0.79x84   &   0.80x59   &   0.80x12   &   0.83x22   &   0.83x85   &   0.83x61   &   0.83x11   &   0.82x54   &   0.81x50    \\
    &        &   0.69x61   &   0.76x89   &   0.77x63   &   0.77x19   &   0.80x02   &   0.80x64   &   0.80x43   &   0.79x89(1)&   0.79x36   &   0.78x34    \\[2pt]
115 & 6.088  &   0.94x68   &   1.02x92   &   1.03x14   &   1.02x12   &   1.06x68   &   1.06x59   &   1.05x75   &   1.05x13   &   1.03x88   &   1.02x22    \\
    &        &   0.88x81   &   0.96x56   &   0.96x81   &   0.95x87   &   0.99x83   &   0.99x77   &   0.99x02   &   0.98x34(1)&   0.97x19(1)&   0.95x61    \\[2pt]
120 & 6.175  &   1.29x26   &   1.38x16   &   1.37x26   &   1.35x22   &   1.42x19   &   1.40x74   &   1.38x81   &   1.37x98   &   1.35x57   &   1.32x86    \\
    &        &   1.15x59   &   1.23x52   &   1.22x80   &   1.21x02   &   1.26x58   &   1.25x32(1)&   1.23x69(1)&   1.22x75(2)&   1.20x64(2)&   1.18x19    \\[2pt]
\end{tabular}
\end{ruledtabular}
\end{table*}
\endgroup

%
%
%
\begingroup
\squeezetable
\begin{table*}
\caption{Self-energy correction for $np_{3/2}$ states.
\label{tab:p3}}
\begin{ruledtabular}
\begin{tabular}{ll..........}
 $Z$ & $R$     &\multicolumn{1}{c}{(2,2)}       &\multicolumn{1}{c}{(2,3)}       &\multicolumn{1}{c}{(2,4)}       &\multicolumn{1}{c}{(2,5)}       &\multicolumn{1}{c}{(3,3)}       &\multicolumn{1}{c}{(3,4)}       &\multicolumn{1}{c}{(3,5)}       &\multicolumn{1}{c}{(4,4)}       &\multicolumn{1}{c}{(4,5)}       &\multicolumn{1}{c}{(5,5)}     \\ \colrule\\[-5pt]
 10 & 3.005  &   0.13x04   &   0.13x35   &   0.13x16   &   0.13x01   &   0.14x21   &   0.14x54   &   0.14x57   &   0.14x74   &   0.14x96   &   0.15x03    \\[2pt]
 15 & 3.189  &   0.13x66   &   0.13x99   &   0.13x80   &   0.13x65   &   0.14x90   &   0.15x25   &   0.15x27   &   0.15x46   &   0.15x69   &   0.15x76    \\[2pt]
 20 & 3.476  &   0.14x38   &   0.14x73   &   0.14x55   &   0.14x40   &   0.15x72   &   0.16x07   &   0.16x10   &   0.16x30   &   0.16x54   &   0.16x62    \\[2pt]
 25 & 3.706  &   0.15x19   &   0.15x56   &   0.15x37   &   0.15x23   &   0.16x63   &   0.16x99   &   0.17x02   &   0.17x25   &   0.17x49   &   0.17x58    \\[2pt]
 30 & 3.929  &   0.16x06   &   0.16x46   &   0.16x27   &   0.16x12   &   0.17x61   &   0.17x99   &   0.18x03   &   0.18x27   &   0.18x52   &   0.18x61    \\[2pt]
 35 & 4.163  &   0.16x99   &   0.17x41   &   0.17x22   &   0.17x06   &   0.18x66   &   0.19x06   &   0.19x09   &   0.19x36   &   0.19x62   &   0.19x72    \\[2pt]
 40 & 4.270  &   0.17x96   &   0.18x41   &   0.18x21   &   0.18x05   &   0.19x77   &   0.20x19   &   0.20x22   &   0.20x52   &   0.20x78   &   0.20x90    \\[2pt]
 45 & 4.494  &   0.18x97   &   0.19x45   &   0.19x25   &   0.19x08   &   0.20x93   &   0.21x37   &   0.21x40   &   0.21x73   &   0.22x00   &   0.22x13    \\[2pt]
 50 & 4.654  &   0.20x01   &   0.20x53   &   0.20x32   &   0.20x15   &   0.22x14   &   0.22x60   &   0.22x64   &   0.22x99   &   0.23x28   &   0.23x41    \\[2pt]
 55 & 4.804  &   0.21x07   &   0.21x65   &   0.21x43   &   0.21x24   &   0.23x40   &   0.23x88   &   0.23x92   &   0.24x31   &   0.24x61   &   0.24x75    \\[2pt]
 60 & 4.912  &   0.22x16   &   0.22x79   &   0.22x56   &   0.22x37   &   0.24x71   &   0.25x21   &   0.25x25   &   0.25x68   &   0.25x99   &   0.26x15    \\[2pt]
 65 & 5.060  &   0.23x28   &   0.23x97   &   0.23x73   &   0.23x53   &   0.26x05   &   0.26x60   &   0.26x63   &   0.27x10   &   0.27x43   &   0.27x60    \\[2pt]
 70 & 5.311  &   0.24x41   &   0.25x17   &   0.24x93   &   0.24x72   &   0.27x45   &   0.28x03   &   0.28x07   &   0.28x58   &   0.28x92   &   0.29x10    \\[2pt]
 75 & 5.339  &   0.25x56   &   0.26x41   &   0.26x16   &   0.25x94   &   0.28x89   &   0.29x51   &   0.29x55   &   0.30x11   &   0.30x47   &   0.30x67    \\
    &        &   0.25x55   &   0.26x40   &   0.26x15   &   0.25x93   &   0.28x88   &   0.29x50   &   0.29x54   &   0.30x10   &   0.30x46   &   0.30x66    \\[2pt]
 80 & 5.463  &   0.26x72   &   0.27x67   &   0.27x42   &   0.27x18   &   0.30x38   &   0.31x05   &   0.31x09   &   0.31x71   &   0.32x08   &   0.32x30    \\
    &        &   0.26x71   &   0.27x66   &   0.27x41   &   0.27x17   &   0.30x36   &   0.31x03   &   0.31x07   &   0.31x69   &   0.32x07   &   0.32x28    \\[2pt]
 85 & 5.539  &   0.27x89   &   0.28x96   &   0.28x71   &   0.28x46   &   0.31x91   &   0.32x64   &   0.32x69   &   0.33x36   &   0.33x76   &   0.33x99    \\
    &        &   0.27x87   &   0.28x94   &   0.28x69   &   0.28x44   &   0.31x89   &   0.32x61   &   0.32x66   &   0.33x33   &   0.33x73   &   0.33x96    \\[2pt]
 90 & 5.710  &   0.29x07   &   0.30x28   &   0.30x03   &   0.29x77   &   0.33x50   &   0.34x28   &   0.34x34   &   0.35x07   &   0.35x50   &   0.35x75    \\
    &        &   0.29x04   &   0.30x24   &   0.30x00   &   0.29x73   &   0.33x46   &   0.34x24   &   0.34x30   &   0.35x03   &   0.35x45   &   0.35x70    \\[2pt]
 95 & 5.905  &   0.30x24   &   0.31x61   &   0.31x38   &   0.31x11   &   0.35x12   &   0.35x98   &   0.36x05   &   0.36x85   &   0.37x30   &   0.37x57    \\
    &        &   0.30x20   &   0.31x56   &   0.31x32   &   0.31x05   &   0.35x06   &   0.35x92   &   0.35x98   &   0.36x78   &   0.37x23   &   0.37x50    \\[2pt]
100 & 5.857  &   0.31x41   &   0.32x97   &   0.32x75   &   0.32x47   &   0.36x79   &   0.37x73   &   0.37x81   &   0.38x68   &   0.39x17   &   0.39x46    \\
    &        &   0.31x35   &   0.32x89   &   0.32x67   &   0.32x39   &   0.36x70   &   0.37x64   &   0.37x71   &   0.38x58   &   0.39x06   &   0.39x35    \\[2pt]
105 & 5.919  &   0.32x56   &   0.34x33   &   0.34x14   &   0.33x85   &   0.38x49   &   0.39x53   &   0.39x63   &   0.40x56   &   0.41x09   &   0.41x40    \\
    &        &   0.32x48(1)&   0.34x21   &   0.34x02   &   0.33x73   &   0.38x36   &   0.39x38   &   0.39x48   &   0.40x41   &   0.40x93   &   0.41x24    \\[2pt]
110 & 5.993  &   0.33x68   &   0.35x67   &   0.35x52   &   0.35x23   &   0.40x20   &   0.41x34   &   0.41x46   &   0.42x47   &   0.43x04   &   0.43x38    \\
    &        &   0.33x56(1)&   0.35x52   &   0.35x35   &   0.35x06   &   0.40x02   &   0.41x13   &   0.41x25   &   0.42x25   &   0.42x81   &   0.43x15    \\[2pt]
115 & 6.088  &   0.34x73   &   0.36x98   &   0.36x88   &   0.36x60   &   0.41x89   &   0.43x14   &   0.43x29   &   0.44x37   &   0.44x98   &   0.45x35(1) \\
    &        &   0.34x57(1)   &   0.36x76   &   0.36x65   &   0.36x35   &   0.41x63   &   0.42x85   &   0.43x00   &   0.44x07   &   0.44x67   &   0.45x02(1) \\[2pt]
120 & 6.175  &   0.35x67   &   0.38x19   &   0.38x16   &   0.37x89   &   0.43x48   &   0.44x84   &   0.45x04   &   0.46x17   &   0.46x83   &   0.47x22(1) \\
    &        &   0.35x48(1)   &   0.37x92   &   0.37x86   &   0.37x57   &   0.43x16(1)&   0.44x48   &   0.44x66   &   0.45x80   &   0.46x43   &   0.46x81(1) \\[2pt]
\end{tabular}
\end{ruledtabular}
\end{table*}
\endgroup

%
%
%
\begingroup
\squeezetable
\begin{table*}
\caption{Self-energy correction for $nd_{3/2}$ states.
\label{tab:d3}}
\begin{ruledtabular}
\begin{tabular}{ll..........}
 $Z$ & $R$     &\multicolumn{1}{c}{(3,3)}       &\multicolumn{1}{c}{(3,4)}       &\multicolumn{1}{c}{(3,5)}       &\multicolumn{1}{c}{(4,4)}       &\multicolumn{1}{c}{(4,5)}       &\multicolumn{1}{c}{(5,5)}     \\ \colrule\\[-5pt]
 10 & 3.005  &  -0.04x27(1)&  -0.03x52   &  -0.03x44(1)&  -0.04x07   &  -0.03x71   &  -0.03x95(1) \\[2pt]
 15 & 3.189  &  -0.04x24   &  -0.03x49   &  -0.03x41   &  -0.04x03   &  -0.03x68   &  -0.03x91    \\[2pt]
 20 & 3.476  &  -0.04x20   &  -0.03x45   &  -0.03x37   &  -0.03x99   &  -0.03x63   &  -0.03x87    \\[2pt]
 25 & 3.706  &  -0.04x15   &  -0.03x40   &  -0.03x32   &  -0.03x93   &  -0.03x58   &  -0.03x81    \\[2pt]
 30 & 3.929  &  -0.04x10   &  -0.03x34   &  -0.03x27   &  -0.03x87   &  -0.03x51   &  -0.03x74    \\[2pt]
 35 & 4.163  &  -0.04x04   &  -0.03x28   &  -0.03x20   &  -0.03x79   &  -0.03x43   &  -0.03x66    \\[2pt]
 40 & 4.270  &  -0.03x96   &  -0.03x20   &  -0.03x13   &  -0.03x71   &  -0.03x34   &  -0.03x56    \\[2pt]
 45 & 4.494  &  -0.03x88   &  -0.03x10   &  -0.03x04   &  -0.03x60   &  -0.03x23   &  -0.03x45    \\[2pt]
 50 & 4.654  &  -0.03x78   &  -0.03x00   &  -0.02x93   &  -0.03x48   &  -0.03x10   &  -0.03x31    \\[2pt]
 55 & 4.804  &  -0.03x66   &  -0.02x87   &  -0.02x80   &  -0.03x34   &  -0.02x95   &  -0.03x16    \\[2pt]
 60 & 4.912  &  -0.03x53   &  -0.02x72   &  -0.02x66   &  -0.03x17   &  -0.02x77   &  -0.02x98    \\[2pt]
 65 & 5.060  &  -0.03x38   &  -0.02x55   &  -0.02x49   &  -0.02x98   &  -0.02x57   &  -0.02x76    \\[2pt]
 70 & 5.311  &  -0.03x21   &  -0.02x36   &  -0.02x29   &  -0.02x76   &  -0.02x33   &  -0.02x52    \\[2pt]
 75 & 5.339  &  -0.03x02   &  -0.02x13   &  -0.02x06   &  -0.02x51   &  -0.02x06   &  -0.02x23    \\[2pt]
 80 & 5.463  &  -0.02x79   &  -0.01x87   &  -0.01x80   &  -0.02x21   &  -0.01x74   &  -0.01x90    \\[2pt]
 85 & 5.539  &  -0.02x54   &  -0.01x57   &  -0.01x49   &  -0.01x88   &  -0.01x37   &  -0.01x52    \\[2pt]
 90 & 5.710  &  -0.02x25   &  -0.01x23   &  -0.01x14   &  -0.01x49   &  -0.00x95   &  -0.01x08    \\[2pt]
 95 & 5.905  &  -0.01x92   &  -0.00x83   &  -0.00x73   &  -0.01x04   &  -0.00x46   &  -0.00x58    \\[2pt]
100 & 5.857  &  -0.01x54   &  -0.00x37   &  -0.00x26   &  -0.00x53   &   0.00x10   &   0.00x01    \\[2pt]
105 & 5.919  &  -0.01x11   &   0.00x15   &   0.00x29   &   0.00x06   &   0.00x74   &   0.00x68    \\
    &        &  -0.01x12   &   0.00x14   &   0.00x28   &   0.00x05   &   0.00x73   &   0.00x67    \\[2pt]
110 & 5.993  &  -0.00x62   &   0.00x74   &   0.00x91   &   0.00x74   &   0.01x48   &   0.01x45    \\
    &        &  -0.00x63   &   0.00x73   &   0.00x90   &   0.00x72   &   0.01x46   &   0.01x43    \\[2pt]
115 & 6.088  &  -0.00x07   &   0.01x42   &   0.01x63   &   0.01x51   &   0.02x32   &   0.02x32    \\
    &        &  -0.00x09   &   0.01x39   &   0.01x60   &   0.01x48   &   0.02x29   &   0.02x29    \\[2pt]
120 & 6.175  &   0.00x55   &   0.02x19   &   0.02x45   &   0.02x38   &   0.03x28   &   0.03x32(1) \\
    &        &   0.00x51(1)&   0.02x14   &   0.02x39   &   0.02x32   &   0.03x21   &   0.03x24(1) \\[2pt]
\end{tabular}
\end{ruledtabular}
\end{table*}
\endgroup

%
%
%
\begingroup
\squeezetable
\begin{table*}
\caption{Self-energy correction for $nd_{5/2}$ states.
\label{tab:d5}}
\begin{ruledtabular}
\begin{tabular}{ll..........}
 $Z$ & $R$     &\multicolumn{1}{c}{(3,3)}       &\multicolumn{1}{c}{(3,4)}       &\multicolumn{1}{c}{(3,5)}       &\multicolumn{1}{c}{(4,4)}       &\multicolumn{1}{c}{(4,5)}       &\multicolumn{1}{c}{(5,5)}     \\ \colrule\\[-5pt]
 10 & 3.005  &   0.04x08   &   0.03x77   &   0.03x54   &   0.04x28   &   0.04x24   &   0.04x40    \\[2pt]
 15 & 3.189  &   0.04x12   &   0.03x81   &   0.03x58   &   0.04x33   &   0.04x29   &   0.04x45    \\[2pt]
 20 & 3.476  &   0.04x17   &   0.03x86   &   0.03x63   &   0.04x40   &   0.04x35   &   0.04x52    \\[2pt]
 25 & 3.706  &   0.04x24   &   0.03x93   &   0.03x70   &   0.04x48   &   0.04x43   &   0.04x60    \\[2pt]
 30 & 3.929  &   0.04x32   &   0.04x01   &   0.03x77   &   0.04x57   &   0.04x53   &   0.04x70    \\[2pt]
 35 & 4.163  &   0.04x41   &   0.04x10   &   0.03x86   &   0.04x67   &   0.04x63   &   0.04x81    \\[2pt]
 40 & 4.270  &   0.04x52   &   0.04x20   &   0.03x96   &   0.04x79   &   0.04x75   &   0.04x94    \\[2pt]
 45 & 4.494  &   0.04x63   &   0.04x31   &   0.04x07   &   0.04x93   &   0.04x89   &   0.05x08    \\[2pt]
 50 & 4.654  &   0.04x75   &   0.04x43   &   0.04x19   &   0.05x07   &   0.05x04   &   0.05x24    \\[2pt]
 55 & 4.804  &   0.04x89   &   0.04x57   &   0.04x32   &   0.05x23   &   0.05x20   &   0.05x42    \\[2pt]
 60 & 4.912  &   0.05x03   &   0.04x72   &   0.04x46   &   0.05x41   &   0.05x38   &   0.05x60    \\[2pt]
 65 & 5.060  &   0.05x19   &   0.04x87   &   0.04x62   &   0.05x60   &   0.05x57   &   0.05x81    \\[2pt]
 70 & 5.311  &   0.05x36   &   0.05x04   &   0.04x78   &   0.05x80   &   0.05x78   &   0.06x02    \\[2pt]
 75 & 5.339  &   0.05x53   &   0.05x22   &   0.04x96   &   0.06x01   &   0.06x00   &   0.06x26    \\[2pt]
 80 & 5.463  &   0.05x72   &   0.05x41   &   0.05x14   &   0.06x24   &   0.06x23   &   0.06x50    \\[2pt]
 85 & 5.539  &   0.05x91   &   0.05x61   &   0.05x34   &   0.06x48   &   0.06x47   &   0.06x76    \\[2pt]
 90 & 5.710  &   0.06x12   &   0.05x82   &   0.05x54   &   0.06x73   &   0.06x73   &   0.07x04    \\[2pt]
 95 & 5.905  &   0.06x33   &   0.06x03   &   0.05x75   &   0.07x00   &   0.07x00   &   0.07x32    \\[2pt]
100 & 5.857  &   0.06x54   &   0.06x26   &   0.05x97   &   0.07x27   &   0.07x28   &   0.07x62    \\[2pt]
105 & 5.919  &   0.06x77   &   0.06x48   &   0.06x19   &   0.07x55   &   0.07x57   &   0.07x92    \\[2pt]
110 & 5.993  &   0.06x99   &   0.06x72   &   0.06x42   &   0.07x83   &   0.07x86   &   0.08x24    \\[2pt]
115 & 6.088  &   0.07x22   &   0.06x95   &   0.06x65   &   0.08x12   &   0.08x16   &   0.08x55(1) \\[2pt]
120 & 6.175  &   0.07x45   &   0.07x19   &   0.06x88   &   0.08x41   &   0.08x46   &   0.08x87(1) \\[2pt]
\end{tabular}
\end{ruledtabular}
\end{table*}
\endgroup

\begin{table}[ht]
\begin{center}
\vspace{1mm}
\caption{ 
The self energy function $F(\alpha Z)$, defined
 by $\Delta E^{\rm SE}=\frac{\alpha}{\pi}\frac{(\alpha Z)^4}{n^3}F(\alpha Z)mc^2$,
 for H-like ions. $\la v|h^{\rm SE}|v\ra$ denotes the results of the
 model SE operator approach, $\la v|V_{\rm loc}^{\rm SE}|v\ra$ is the contribution of
 the local part of the model SE operator, 
and "Exact" labels the results of the exact calculation.
}
\label{tab:H-like}
\vspace{2mm}

\begin{tabular}{ccccc}
\hline \hline\\[-2mm]
$Z$ & State &   $\la v|V_{\rm loc}^{\rm SE}|v\ra$ &  $\la v|h^{\rm SE}|v\ra$  & Exact \\
\hline\\[-2mm]
10  & $4s$ &4.60 & 4.96 & 4.97  \\
    & $5s$ &4.59  & 4.96 & 4.99 \\
    & $5p_{1/2}$ &-0.15  & -0.10 & -0.09 \\
    & $5p_{3/2}$ &0.17  & 0.15 & 0.15 \\
    & $5d_{3/2}$ &-0.07  & -0.05 & -0.04 \\
    & $5d_{5/2}$ &0.07  & 0.05 & 0.04 \\
\hline\\[-2mm]
20  & $4s$ &3.11 & 3.57 & 3.58  \\
    & $5s$ &3.11  & 3.57 & 3.59 \\
    & $5p_{1/2}$ &-0.12  & -0.08 & -0.07 \\
    & $5p_{3/2}$ &0.18  & 0.16 & 0.17 \\
    & $5d_{3/2}$ &-0.07  & -0.05 & -0.04 \\
    & $5d_{5/2}$ &0.07  & 0.05 & 0.05 \\
\hline\\[-2mm]
40  & $4s$ &1.91 & 2.51 & 2.52  \\
    & $5s$ &1.90  & 2.50 & 2.52 \\
    & $5p_{1/2}$ &-0.04  & 0.00 & 0.01 \\
    & $5p_{3/2}$ &0.22  & 0.21 & 0.21 \\
    & $5d_{3/2}$ &-0.06  & -0.04 & -0.04 \\
    & $5d_{5/2}$ &0.07  & 0.06 & 0.05 \\
\hline\\[-2mm]
60  & $4s$ &1.46 &2.13 & 2.14  \\
    & $5s$ &1.44  & 2.12 & 2.14 \\
    & $5p_{1/2}$ &0.06  & 0.11 & 0.12 \\
    & $5p_{3/2}$ &0.26  & 0.26 & 0.26 \\
    & $5d_{3/2}$ &-0.05  & -0.04 & -0.03 \\
    & $5d_{5/2}$ &0.08  & 0.06 & 0.06 \\
\hline\\[-2mm]
83  & $4s$ &1.37 & 2.08 & 2.09  \\
    & $5s$ &1.34  & 2.05 & 2.06 \\
    & $5p_{1/2}$ &0.21  & 0.29 & 0.30 \\
    & $5p_{3/2}$ &0.30  & 0.33 & 0.33 \\
    & $5d_{3/2}$ &-0.04  & -0.03 & -0.02 \\
    & $5d_{5/2}$ &0.09  & 0.07 & 0.07 \\
\hline\\[-2mm]
92  & $4s$ &1.44 & 2.15 & 2.16  \\
    & $5s$ &1.40  & 2.10 & 2.12 \\
    & $5p_{1/2}$ &0.29  & 0.40 & 0.41 \\
    & $5p_{3/2}$ &0.32  & 0.36 & 0.36 \\
    & $5d_{3/2}$ &-0.03  & -0.02 & -0.01 \\
    & $5d_{5/2}$ &0.09  & 0.08 & 0.07 \\
\hline\\[-2mm]
\hline
\end{tabular}
\end{center}
\end{table}
\begin{table}[ht]
\begin{center}
\vspace{1mm}
\caption{ The self energy function $F(\alpha Z)$, defined
 by $\Delta E^{\rm SE}=\frac{\alpha}{\pi}\frac{(\alpha Z)^4}{n^3}F(\alpha Z)mc^2$,
 for neutral alkali metals
in different potentials.  
$\la v|h^{\rm SE}|v\ra$ denotes the results of the
 model operator approach, calculated by averaging the model
 SE operator with the valence electron wave
 function in the corresponding potential.
$\la v|V_{\rm loc}^{\rm SE}|v\ra$ is the contribution of
 the local part of the model SE operator, 
and "Exact" labels the results of the
 exact calculation of Ref. \cite{sap02}.
}
\label{tab:alkali}
\vspace{2mm}

\begin{tabular}{cccccc}
\hline \hline
&&&&&\\[-2mm]
Atom & Method  & \hspace{1mm}  $x_{\alpha}=0$   \hspace{1mm}
&   \hspace{1mm}  $x_{\alpha}=1/3$ \hspace{1mm}
&   \hspace{1mm}  $x_{\alpha}=2/3$ \hspace{1mm}
&   \hspace{1mm}  $x_{\alpha}=1$ \hspace{1mm}
   \\[0mm]
\hline
&&&&& \\[-1mm]
Na $3s_{1/2}$ & $\la v|V_{\rm loc}^{\rm SE}|v\ra$ &   0.166~ &   0.163~ &  0.176~  & 0.214~  \\[0mm]
              &  $\la v|h^{\rm SE}|v\ra$    &   0.170~ &   0.168~ &  0.183~  & 0.224~ \\[0mm]
              & Exact \cite{sap02}   & 0.169 & 0.167 & 0.181  & 0.223
               \\[2mm]
K  $4s_{1/2}$ &  $\la v|V_{\rm loc}^{\rm SE}|v\ra$  & 0.067~ & 0.067~  & 0.076~  & 0.100~  \\[0mm]
              &  $\la v|h^{\rm SE}|v\ra$  & 0.072~ & 0.072~  & 0.083~  & 0.110~  \\[0mm]
              &    Exact  \cite{sap02}    & 0.072 & 0.072 & 0.083 & 0.110
               \\[2mm]
Rb $5s_{1/2}$ &  $\la v|V_{\rm loc}^{\rm SE}|v\ra$ & 0.0187~ &  0.0193~  &  0.0230~  & 0.0320~
              \\[0mm]
              &  $\la v|h^{\rm SE}|v\ra$  & 0.0229~ &  0.0237~  &  0.0284~  & 0.0397~
               \\[0mm]
              &  Exact \cite{sap02}    & 0.0228 & 0.0236 & 0.0283 & 0.0396
               \\[2mm]
Cs $6s_{1/2}$ &  $\la v|V_{\rm loc}^{\rm SE}|v\ra$ &  0.0093~   & 0.0097~    & 0.0118~    & 0.0171~
               \\[0mm]
              &  $\la v|h^{\rm SE}|v\ra$  &  0.0127~   & 0.0132~    & 0.0163~    & 0.0236~
               \\[0mm]
              &  Exact \cite{sap02}    & 0.0126 & 0.0132 & 0.0162 & 0.0235
              \\[2mm]
Fr $7s_{1/2}$ &  $\la v|V_{\rm loc}^{\rm SE}|v\ra$ & 0.0047~ &   0.0052~  & 0.0067~ & 0.0102~ \\[0mm]
              &  $\la v|h^{\rm SE}|v\ra$  & 0.0069~ &   0.0076~  & 0.0099~ & 0.0151~  \\[0mm]
              &  Exact \cite{sap02}    & 0.0068 & 0.0075 & 0.0098 & 0.0150
              \\[2mm]
\hline
\hline
\end{tabular}
\end{center}
\end{table}
%
\begin{table}[ht]
\begin{center}
\vspace{1mm}
\caption{The self energy contribution to the $4s-4p_{1/2}$, $4s-4p_{3/2}$,
 $4p_{1/2}-4d_{3/2}$, $4p_{3/2}-4d_{3/2}$, and $4p_{3/2}-4d_{5/2}$
transition energies
in Cu-like ions, in eV.
$\la v|h^{\rm SE}|v\ra$ denotes the results of the
 model operator approach, calculated by averaging the
 model SE operator with the Dirac-Kohn-Sham wave
function of the valence electron.
"Exact" labels the results of the exact calculation of Ref. \cite{che06}.
}
\label{tab:cu}
\vspace{2mm}

\begin{tabular}{cccc}
\hline \hline\\[-2mm]
Ion & Transition &  $\la v|h^{\rm SE}|v\ra$  & Exact \cite{che06} \\
\hline\\[-2mm]
Yb$^{41+}$  & $4s-4p_{1/2}$  & -1.29 & -1.28  \\
            & $4s-4p_{3/2}$  & -1.21 & -1.21 \\
            & $4p_{1/2}-4d_{3/2}$  & -0.10 & -0.11 \\
            & $4p_{3/2}-4d_{3/2}$  & -0.18 & -0.18 \\
            & $4p_{3/2}-4d_{5/2}$  & -0.14 & -0.14 \\
\hline\\[-1.5mm]
W$^{45+}$  & $4s-4p_{1/2}$  & -1.64 &-1.64  \\
           & $4s-4p_{3/2}$  & -1.55 & -1.56 \\
            & $4p_{1/2}-4d_{3/2}$  & -0.16 & -0.17 \\
            & $4p_{3/2}-4d_{3/2}$  & -0.25 & -0.25 \\
            & $4p_{3/2}-4d_{5/2}$  & -0.19 & -0.19 \\

\hline\\[-1.5mm]
Os$^{47+}$  & $4s-4p_{1/2}$  & -1.85 &-1.84  \\
            & $4s-4p_{3/2}$  & -1.75 & -1.76 \\
            & $4p_{1/2}-4d_{3/2}$  & -0.19 & -0.20 \\
            & $4p_{3/2}-4d_{3/2}$  & -0.28 & -0.28 \\
            & $4p_{3/2}-4d_{5/2}$  & -0.22 & -0.22 \\

\hline\\[-1.5mm]
Au$^{50+}$  & $4s-4p_{1/2}$  & -2.18 &-2.18  \\
            & $4s-4p_{3/2}$  & -2.10 & -2.10 \\
            & $4p_{1/2}-4d_{3/2}$  & -0.26 & -0.28 \\
            & $4p_{3/2}-4d_{3/2}$  & -0.35 & -0.35 \\
            & $4p_{3/2}-4d_{5/2}$  & -0.27 & -0.28 \\
\hline\\[-1.5mm]
Pb$^{53+}$  & $4s-4p_{1/2}$  & -2.57 &-2.56  \\
            & $4s-4p_{3/2}$  & -2.49 & -2.50 \\
            & $4p_{1/2}-4d_{3/2}$  & -0.35 & -0.37 \\
            & $4p_{3/2}-4d_{3/2}$  & -0.43 & -0.43 \\
            & $4p_{3/2}-4d_{5/2}$  & -0.34 & -0.34 \\

\hline\\[-1.5mm]
Bi$^{54+}$  & $4s-4p_{1/2}$  & -2.71 &-2.70  \\
            & $4s-4p_{3/2}$  & -2.64 & -2.64 \\
            & $4p_{1/2}-4d_{3/2}$  & -0.39 & -0.40 \\
            & $4p_{3/2}-4d_{3/2}$  & -0.46 & -0.46 \\
            & $4p_{3/2}-4d_{5/2}$  & -0.36 & -0.37 \\

\hline\\[-1.5mm]
Th$^{61+}$  & $4s-4p_{1/2}$  & -3.85 &-3.85  \\
            & $4s-4p_{3/2}$  & -3.88 & -3.89 \\
            & $4p_{1/2}-4d_{3/2}$  & -0.73 & -0.74 \\
            & $4p_{3/2}-4d_{3/2}$  & -0.70 & -0.71 \\
            & $4p_{3/2}-4d_{5/2}$  & -0.56 & -0.57 \\

\hline\\[-1.5mm]
U$^{63+}$  & $4s-4p_{1/2}$  & -4.24 &-4.24  \\
           & $4s-4p_{3/2}$  & -4.32 & -4.33 \\
           & $4p_{1/2}-4d_{3/2}$  & -0.87 & -0.88 \\
            & $4p_{3/2}-4d_{3/2}$  & -0.79 & -0.79 \\
            & $4p_{3/2}-4d_{5/2}$  & -0.63 & -0.65 \\
\hline
\end{tabular}
\end{center}
\end{table}

\begin{table*}[ht]
\begin{center}
\vspace{1mm}
\caption{The self energy contribution to the binding energy of the valence
electrons in Rg, Cn, E119 and E120, in eV.
In this work, the perturbation theory (PT) value is obtained by
averaging the model SE operator with the Dirac-Fock wave
function of the valence electron, while the DF and total DF
values are obtained by including
this operator into the DF equations.
The DF value is given by the SE contribution to 
the one-electron binding energy 
  whereas the total DF value is obtained as 
the difference between the SE contributions to 
the total DF energies of the atom and the ion
 (the  $^2S_{1/2}\rightarrow ^1S_0$ 
and the $^1S_{0}\rightarrow ^2S_{1/2}$
transitions are considered for Rg and Cn, respectively;
see the text of the paper).
 In Refs. \cite{lab99,goi09}, the calculations
were performed with a local DF potential. 
}
\label{tab:comp}
\vspace{2mm}

\begin{tabular}{cccccc}
\hline \hline
Atom & Valence electron & Method & This work & Ref. \cite{goi09} & Other works \\
\hline
Rg   & $7s$  & PT & -0.088 & -0.089 & -0.087$^a$\\
     &        & DF & -0.105 & -0.102 &  \\
     &        & Total DF & -0.096 &       \\
     &        & Welton method &  &  &  -0.084$^b$   \\
     &        & Local SE pot. &  & &  -0.089$^c$ \\
\hline
Cn   & $7s$  & PT & -0.101 & -0.103 &   \\
     &       & DF & -0.105 & -0.110 & \\
     &       & Total DF & -0.098 &       \\
     &        & Local SE pot. &  & &  -0.091$^c$ \\
\hline
E119   & $8s$ & PT  & -0.0233 & & -0.0274$^a$  \\
       &      & DF  & -0.0250 & &   \\
       &      & Total DF & -0.0232 & &     \\
       &      & Local SE pot. &  & &  -0.0210$^c$ \\
\hline
E120   & $8s$  & PT & -0.0331 &  \\
       &       & DF & -0.0265 &  \\
       &       & Total DF & -0.0250 & &    \\
       &       & Local SE pot. &  & & -0.0226$^c$ \\
\hline
\end{tabular}
\end{center}
$^a$Taken from Ref. \cite{lab99};
$^b$Taken from Ref. \cite{ind07};
$^c$Taken from Ref. \cite{thi10}.\\
\end{table*}

\begin{table}[ht]
\begin{center}
\vspace{1mm}
\caption{The screened self energy for the $2s$, $2p_{1/2}$, and $2p_{3/2}$ states
of Li-like ions, in eV. The Kohn-Sham (KS) and Dirac-Fock (DF) results
 are obtained by calculating the total
ion energy with the model SE operator included into the KS and DF
equations, respectively, and substructing the related energies
evaluated without the
model SE operator and the self energy contributions
 calculated with the hydrogenlike
wave functions.
 Comparison with the calculations performed
by perturbation theory (PT) in the  Kohn-Sham
potential \cite{koz10,sap11}
is given.
}
\label{tab:li}
\vspace{2mm}

\begin{tabular}{cccccc}
\hline \hline
$Z$ & State & KS  &  DF
&   PT \cite{koz10} &PT \cite{sap11}
   \\[0mm]
\hline
20  & \hspace{1mm} $2s$ & -0.047  & -0.045   &   -0.044  & -0.046    \\
    & \hspace{1mm} $2p_{1/2}$ & -0.009 &-0.008           &   -0.008 & -0.008   \\
    &  \hspace{1mm}  $2p_{3/2}$   & -0.012    & -0.011          & -0.013 &  -0.013    \\
\hline
40  & \hspace{1mm} $2s$ & -0.277  & -0.269   &   & -0.260    \\
    & \hspace{1mm} $2p_{1/2}$ & -0.063 &-0.059           &   & -0.059   \\
    &  \hspace{1mm}  $2p_{3/2}$   & -0.077    & -0.073        &  &  -0.085    \\
\hline
50  & \hspace{1mm} $2s$  & -0.50 & -0.49   &-0.48  \\
    & \hspace{1mm} $2p_{1/2}$    & -0.13 & -0.12           &   -0.12      \\
    &  \hspace{1mm}  $2p_{3/2}$   & -0.14       & -0.14          &   -0.16    \\
\hline
60  & \hspace{1mm} $2s$  & -0.84 & -0.82   &  & -0.80 \\
    & \hspace{1mm} $2p_{1/2}$  & -0.24 &    -0.24       &  & -0.25     \\
    &  \hspace{1mm}  $2p_{3/2}$   & -0.24       &-0.23    &      &  -0.27   \\
\hline
74  & \hspace{1mm} $2s$  & -1.58 &-1.55 &   &  -1.55  \\
    & \hspace{1mm} $2p_{1/2}$  & -0.57 &-0.56    &       &  -0.62      \\
    &  \hspace{1mm}  $2p_{3/2}$   & -0.45       & -0.43   &       & -0.53     \\
\hline
83  & \hspace{1mm} $2s$ & -2.35  & -2.25   &   -2.32  &   -2.26   \\
    & \hspace{1mm} $2p_{1/2}$       & -0.97   &-0.98           &     -1.07  &  -1.07    \\
    &  \hspace{1mm}  $2p_{3/2}$      &    -0.65   & -0.61          &  -0.75  &  -0.76 \\
\hline
92  & \hspace{1mm} $2s$ & -3.47  & -3.19   &     &   -3.81   \\
    & \hspace{1mm} $2p_{1/2}$       & -1.67   &-1.69           &    &  -1.58    \\
    &  \hspace{1mm}  $2p_{3/2}$      &    -0.91   & -0.86          &   &  -1.04 \\

\hline
\hline
\end{tabular}
\end{center}
\end{table}
%




\end{document}